\magnification=1200
\overfullrule=0pt
\baselineskip=20pt
\parskip=0pt
\def\dag{\dagger}
\def\del{\partial}

\def\a{\alpha}             
\def\b{\beta}              
        
\def\d{\delta}        
\def\e{{\rm e}}           
\def\z{\zeta}              
\def\j{\eta}

\def\m{\mu}	           
\def\n{\nu}                
\def\x{\xi}              
                  
\def\p{\pi}        \def\P{\Pi}      
\def\r{\rho}

\def\y{\psi}

\def\yh{\hat{\y}}
\def\yhd{\hat{\y}^{\dag}}

\def\w{\omega}     
   
\def\br{\langle}

\def\ve{\vert}

\def\zbar{\bar{z}}

\def\eqref#1{(\ref{#1})}

{\settabs 5 \columns
\+&&&&UdeM-GPP-TH-97-42\cr
\+&&&&UB-ECM-PF 97/17\cr}
\bigskip
\centerline{\bf FIELD THEORETICAL APPROACH TO QUANTUM HALL FERROMAGNETS}
\bigskip
\bigskip
\centerline{Rashmi Ray$^1$}
\centerline{Laboratoire de Physique Nucl\'eaire}
\centerline{Universit\'e de Montr\'eal}
\centerline{Montr\'eal, Quebec H3C 3J7, Canada}
\bigskip
\centerline{Joan Soto$^2$}
\centerline{Departament d'Estructura i Constituents de la Mat\`eria }
\centerline{ Facultat de Fisica}
\centerline{Universitat de Barcelona}
\centerline{Diagonal, 647}
\centerline{E-08028-Barcelona-Catalonia-Spain}
\bigskip
\magnification = 1200
\centerline {\bf Abstract}
\bigskip
We present a quantum field theoretical analysis of a $\n = 1$ quantum
Hall system when the effective Land\'e $g$ factor is small. We clearly
demonstrate that the ground state of the system is ferromagnetic. We
note that it is the short range component of the Coulomb interaction
which is instrumental in aligning the spins. We then go on to derive
the effective lagrangian for the lowest lying spin excitations. At the
leading order, apart from the usual O(3) sigma model terms, we find
a term proportional to the topological Pontryagin density and a long
range Coulomb interaction term between these densities. Beyond the
leading order in the derivative expansion, we find an interesting
Chern-Simons term constructed out of the basic spin variables. For
low enough energies, however, we notice that the effects of mixing of
higher Landau levels is more important than the next to leading terms in
the derivative expansion. We provide
a systematic way of calculating these corrections.
\bigskip
\noindent{$^1$ rray@lpshpb.lps.umontreal.ca}
\noindent{$^2$ soto@ecm.ub.es}
\vfill
\eject
\magnification=1200
\overfullrule=0pt
\baselineskip=20pt
\parskip=0pt
\def\dag{\dagger}
\def\del{\partial}

\def\a{\alpha}             
\def\b{\beta}              
        
\def\d{\delta}        
\def\e{{\rm e}}           
\def\z{\zeta}              
\def\j{\eta}

\def\m{\mu}	           
\def\n{\nu}                
\def\x{\xi}              
                  
\def\p{\pi}        \def\P{\Pi}      
\def\r{\rho}

\def\y{\psi}

\def\yh{\hat{\y}}
\def\yhd{\hat{\y}^{\dag}}

\def\w{\omega}     
   
\def\br{\langle}
\def\ket{\rangle}
\def\ve{\vert}

\def\zbar{\bar{z}}

\def\eqref#1{(\ref{#1})}
\def\zexp{\e^{-{{B}\over 2}\ve z \ve^2 }}

\def\zpexp{\e^{-{{B}\over 2}\ve \zp \ve^2 }}

\def\zp{z^{\prime }}
\def\zbarp{\bar{\zp }}
\def\zzprint{\int d^2z d^2\zp }
\def\zzprexp{\zzprint \e^{-{{B}\over 2}(\vert z \vert^2 +\vert \zp \vert^2 )}}

\def\dt{\int dt}
\def\dx{\int d\vec x}
\def\dxx{\int d\vec x^{\prime } }
\def\dxi{\int d^2 \x \ \e^{-{{B}\over 2}{\vert \x \vert^2}}}

\def\Phx{\hat {\P^x }}
\def\Phy{\hat {\P^y }}

\def\Ph{\hat \P }
\def\Phd{\hat {\P^{\dag }}}
\def\Xh{\hat X}
\def\Yh{\hat Y}
\def\zh{\hat z}
\def\zbh{\hat{\zbar }}
\def\Zh{\hat Z}
\def\Zbh{\hat {\bar Z}}

\magnification=1200
{\bf 1. Introduction}
\bigskip
Over the last few years, a lot of interest has been generated over the physics
of the quantum Hall effect of multicomponent systems [1]. These components
could be electron spin, electrons in different valleys (as in Si systems) or
even be electrons in different layers in multi-layered samples.

In this article, we shall focus on the case of a single-layer system, where,
owing to various circumstances, the electron spin becomes a relevant SU(2)
degree of freedom.

Specifically, we shall be discussing systems containing fermions living on a 
plane, interacting via the Coulomb interaction and subjected to a strong
magnetic field orthogonal to the plane. These non-relativistic electrons
couple minimally as well as through the Pauli term to the magnetic field.
In the absence of the Coulomb repulsion, the single-particle spectrum is 
well-known: It comprises of equi-spaced Landau levels, each with an infinite
degeneracy. If the Land\'e $g$ factor of the electrons retains its vacuum 
value, namely, $g=2$, the Zeeman split between opposite spins is precisely
equal to the gap between any two successive Landau levels. Therefore, if
only the lowest Landau level (L.L.L.) is filled, the system is automatically
spin polarised.
In many realistic systems(e.g. GaAs), however, band effects enhance the
Landau level(L.L.) separation by a factor of $ \sim 20 $ and reduce the Zeeman
splitting by $ \sim 4 $ [2]. For typical experimental situations, the L.L.
separation, in temperature units, is $\sim 200K $, the Coulomb interaction is
$\sim 100K$ and the Zeeman split is $\sim 2K $. In view of these numbers, it is 
clear that the effective $g$ factor is $\ll 2$. In this case, the gap between
successive single-particle levels is reduced considerably and the spin
degree of freedom can be expected to play a signifant role in these 
systems.

Also, from the numbers quoted above, it would seem that the relevant gap
in these systems would be the Zeeman gap rather than the Landau gap. However,
in the case of odd-integer fillings, experimentally measured gaps are seen
to exceed the single-particle Zeeman gap by a factor of as much as 20 [3].
Thus it seems that for these odd-integral Hall systems, the interaction
between the electrons plays a decisive role in generating the observed gap.
In these cases, the conventional distinction that is made between the integer
and the fractional Hall effects gets blurred.

Here, we consider the simplest odd-integral Hall system with the filling
fraction $\nu =1$. We find that the incompressible ground state of the system
is spin-polarised. This is true even when $g$ is set equal to zero, when the
underlying microscopic action acquires an exact SU(2) spin symmetry. Thus
the ground state breaks this symmetry down to a U(1) phase symmetry. This
is the hallmark of spontaneous symmetry breaking (SSB) and thus the ground
state is ferromagnetic. The lowest-lying excitations are off course the
neutral gapless magnons which are associated with the broken generators of
the spin SU(2). These are long wavelength spin waves. However, in these
quantum Hall ferromagnets, there are, in contradistinction with usual
ferromagnets, low-lying charged excitations of macroscopic spin and nontrivial
spin order. These are actually the dominant charged excitations whenever
$g \le g_{cr}$ and when $\nu $ deviates slightly from 1 [3]. They are known as
spin-skyrmions and are topological in nature, their topological charge
(Pontryagin number) equalling their electric charge. Under the conditions
just mentioned, they are actually more relevant than charged single-particle
excitations. Experiments [4] seem to provide support for the existence of these
topological excitations. 

The system has already been studied in a variety of ways [3,5,6,7,8]. These works
all convey the essential physics quite accurately. However, it seems to us
that there are lacunae in the actual derivation of the effective action
for the spin starting from a microscopic theory, as we argue below.

In [3,6], the effective action has been motivated rather than derived from
the microscopics. Thus some terms that do not occur in usual ferromagnetic
sigma-model effective actions have not been discussed. The first-quantised
approach espoused in [5] is predicated to discussing only the lowest Landau
level (L.L.L.) and it is not immediately obvious how the effects of L.L.
mixing by the Coulomb term may be incorporated. Furthermore, the distinctness
of the rather different roles played by the long and the short distance
components of the Coulomb interaction in the derivation of the effective
action has not, in our opinion, been suitably emphasized. Namely, that it is
the short distance component that provides the magnon kinetic energy and the
long distance part that mediates a Coulomb repulsion between skyrmions has
not been spelled out clearly enough. The interaction that leads to a 
ferromagnetic ground state is the Coulomb interaction which is somewhat
counter-intuitive as it is a nonlocal interaction. Thus it is in our opinion
vastly clarifying to see ferromagnetism emerge from the short distance part
of this nonlocal interaction. In [8], a Hartree-Fock theory of skyrmions has
been proposed, but the contribution to the effective spin action due to
Landau Level mixing has not been addressed.

The existence of SSB and the dynamics of the resulting Goldstone modes is
studied most easily within a second-quantised (quantum field theoretic)
framework. A further clear advantage of a field theoretic formulation is
that effects due to mixing with the higher L.L. may be systematically 
accounted for. In this article, therefore, we have formulated our problem
in a second-quantised language.

Thus, the system is taken to be governed by a second-quantised fermionic
action whose explicit form will be given in the next section. For it to
desribe a $\nu = 1$ Hall system, the corresponding ground state must be
made up of only the L.L.L. single-particle states. Therefore we make a very
general ansatz for the density matrix describing such a ground state, where
this matrix is constructed out of only the L.L.L. single-particle states.
We then compute the mean energy of the ground state using this ansatz and
the second-quantised Hamiltonian for the system. The requirement that this
energy should be minimised automatically fixes the parameters in the ansatz
such that the matrix describes a ground state with uniform density and
magnetisation. We see that the energy is a minimum when all the spins are
aligned in any given direction, all choices of this direction being 
equivalent. Thus we choose a given direction and thereby break the SU(2)
symmetry of the action spontaneously by our choice of the ground state.

The description of the Goldstone excitations above this ground state is
the clearest in the nonlinear realisation [9] of spontaneous symmetry
breaking. Namely, if a continuous symmetry group $G$ is broken spontaneously
to a group $H$, where $H$ is a proper subgroup of $G$, the Goldstone modes are
described by a slowly-varying unitary matrix U living in the 
coset space $G/H$. U transforms nonlinearly under transformations $g$ belonging
to $G$. The density matrix itself, by construction should transform
covariantly under $g \in G$. In our case we generalise the ansatz to
include the magnons by introducing an appropriate unitary matrix U and
requiring that the resultant density matrix be covariant under SU(2),
\underbar{{\bf \it projected onto the L.L.L.}}. This last requirement 
entails the
inclusion of appropriate derivatives of U to render the matrix covariant.

We further use this density matrix to compute the mean action, which is
now expressed in terms of the unitary matrix U and is the effective action
for the spin degrees of freedom. A rather interesting feature of L.L.L.
projection is that charge and spin densities do not commute. This leads
to the possibility that some non-trivial spin configurations may be charged,
which is realised through the spin-skyrmions.

The effective action thus obtained contains all the usual terms one expects
for the ferromagnetic magnon system. Namely, we get a Wess-Zumino term with a
single time-derivative and a standard magnon kinetic energy term that is
quadratic in spatial derivatives. These ensure that the magnon dispersion
is quadratic, as is the case for ferromagnets. Furthermore, we get the
Coulomb repulsion between the topological charge densities first obtained in
[3]. Over and above these terms, however, we get some terms which 
have been noticed only recently [10]. There is a term proportional to the Pontryagin number
which implies that topological excitations are energetically favoured. Further,
there is a Chern-Simons term that can be constructed out of U which 
may result
in the statistical transmutation of the quasiparticles [11]. These emerge with a
minimum of fuss in our formalism of nonlinear realisation of the spin SU(2).

As mentioned above, another advantage of a field theoretic formulation is the 
ease with which contributions due to mixing with higher L.L. are computed.
Here, we envision a path integral approach, whereby the higher L.L. may be
integrated out explicitly to yield an effective action for the L.L.L.
As an example we show how the magnon kinetic term and the topological term
acquire finite renormalisation due to this mixing.

The organisation of the article is as follows. In section 2, we establish
the notation and describe the projection onto the L.L.L. In section 3, we
show that an ansatz for the density matrix for the ground state involving
only the L.L.L. automatically yields a ferromagnetic ground state when
energy minimisation is required. In section 4, we generalise the ansatz to
include the Goldstone modes and obtain the effective action for the spin
degree of freedom. Leading corrections due to mixing with higher L.L. are
obtained in section 5. We conclude in section 6 with a critical evaluation
of the contents of this article and a discussion of the avenues for further
investigations. Details of the calculation and a brief discussion of nonlinear
realisations of symmetries have been relegated to the appendices.
\bigskip
{\bf 2. Notation and Formulation}
\bigskip
In this section, we shall write down the microscopic action for the system
at hand, solve the single-particle problem in the absence of the Coulomb
interaction and establish the notation. As far as the basic notation is
concerned, we follow [12] as closely as possible.

We also discuss the projection to the L.L.L. and write down the form of
the Coulomb term after projection on to the L.L.L.

The microscopic action for nonrelativistic planar spinning fermions in a
magnetic field perpendicular to the plane is:

$$S=S_0 + S_c \eqno(2.1)$$

where

$$S_0 \equiv \dt \dx \  \yhd (\vec x,t)\bigl[i\del_t -{1\over{2m }}
(\vec p -\vec A )^2 + {{g B}\over{2m }}S_z + \m \bigr ]\yh (\vec x,t)
\eqno(2.2)$$

Here, we have rescaled $e\vec A \rightarrow \vec A $ and thus, $e$ does
not appear explicitly in (2.2). Also,

$$S_c \equiv -{1\over 2}\dt \dx \dxx\  \yhd (\vec x,t)\yhd (\vec x^{\prime },t)
V(\vert \vec x - \vec x^{\prime }\vert )\yh (\vec x^{\prime },t)
\yh (\vec x,t) \eqno(2.3). $$

Here, $m$ is the effective mass, $g$ the effective $g$ factor and
$\m $ the chemical potential that specifies the particle content of the
system. We are going to take $V(\vert \vec x - \vec x^{\prime }\vert )=
{{e^2}\over{\vert \vec x - \vec x^{\prime }\vert }}$, which is the Coulomb
repulsion. Here, we have retained $e$ explicitly, as, in the sequel, this
will be seen to generate a new scale in the problem.

$\yh_{\a }$ is a 2-component fermion operator obeying
$\{ \yh_{\a }(\vec x), \yhd_{\b }(\vec x^{\prime })\} = \d_{\a \b }\d (\vec x -
\vec x^{\prime })$.

Further, let

$$h_0 \equiv {1\over{2m }}(\vec p -\vec A)^2-{{g B}\over {2m }}
S_z \eqno(2.4)$$

and

$$\vec \nabla \wedge \vec A = -B , \eqno(2.5)$$

where $B$ is the applied strong magnetic field.

The single-particle spectrum of $h_0$ is readily obtained.

Let 

$$\Phx \equiv -i\del_x - A^x $$

and

$$\Phy \equiv -i\del_y - A^y \eqno(2.5)$$

where 

$$\vec A = (\a By, -(1-\a )Bx) \eqno(2.6),$$

$\a $ being some arbitrary parameter that interpolates between various
gauge choices and drops out of physical quantities.

Further, let

$$\Ph \equiv \Phx -i \Phy $$

and

$$\Phd \equiv \Phx +i \Phy \eqno(2.7). $$

Then,

$$h_0 = {1\over{2m }}\Phd \Ph + {{B}\over{2m }}(1-g S_z)
\eqno(2.8).$$

The spectrum of $h_0$ is highly degenerate and this degeneracy is exposed
by the ``guiding centre" operator $$\Xh \equiv \hat x - {1\over{B}}\Phy 
\eqno(2.10)$$ which commutes with $h_0$.

The canonical conjugate of $\Xh $ is

$$\Yh \equiv \hat y + {1\over{B}}\Phx \eqno(2.11).$$

We form the holomorphic and anti-holomorphic combinations of these:

$$\Zh \equiv \Xh + i \Yh $$

and

$$\Zbh \equiv \Xh - i \Yh \eqno(2.12).$$

Then,

$$\bigl[ \Zh , \Zbh \bigr] = {2\over{B}} \eqno(2.13).$$

Further,

$$\bigl[ \Zh , \Ph \bigr] = \bigl[ \Zbh , \Ph \bigr] = 0 \eqno(2.14).$$

We choose the eigenstates of $h_0$ to be $\bigl\{ 
\vert n,\x , \a \ket  \bigr\}$
where,

$$\eqalignno{
\Ph \vert n, \x , \a \ket &= \sqrt{2Bn} \vert n-1, \x , \a \ket \cr
\Phd \vert n, \x , \a \ket &= \sqrt{2B(n+1)} \vert n+1, \x , \a \ket \cr
\Zh \vert n, \x , \a \ket &= \x \vert n, \x , \a \ket \cr
\hat{S_z} \vert n,\x , \a \ket &= \a \vert n, \x , \a \ket & (2.15).\cr
}$$

Here, $\a = \pm {1\over 2}$.
Thus,

$$ h_0 \vert n,\x ,\a \ket = E_{n,\a }\vert n,\x ,\a \ket \eqno(2.16)$$

where

$$E_{n,\a }=(n+{1\over2}-{{g \a }\over 2}){\w_c}\eqno(2.17).$$

Here, ${\w_c}\equiv {{B}\over {m}}$, the cyclotron frequency.

Therefore, for $g = 2$, the energy difference between $\a =\pm {1\over 2}$
is precisely ${\w_c}$, as mentioned in the introduction.

Now, $\vert n,\x ,\a \ket = \vert n ,\x \ket \otimes 
\vert \a \ket $, where $\vert \x \ket $ is a coherent state of $\Zh $.

Namely,

$$\Zh \vert \x \ket = \x \vert \x \ket \eqno(2.18).$$

Explicitly,

$$\vert \x \ket = \exp^{{{B}\over 2}\x \Zbh }\vert 0 \ket \eqno(2.19), $$

where $\Zh \vert 0 \ket = 0$.

The resolution of the identity in terms of these coherent states is:

$$I = {{B}\over {2\pi }}\dxi \vert \x \ket \br \bar{\x }\vert \eqno(2.20), $$

where,

$$ d^2\x \equiv d\ (\rm{Re}\x )d\ (\rm{Im}\x ) $$.

The inner product of these coherent states is

$$ \br \bar{\eta }\vert \x \ket = \e^{{{B}\over 2}\bar{\eta }\x }\eqno(2.21).$$

Now, the eigenfunctions are obtained by projecting $\vert n,\x ,\a \ket $ 
onto coordinate space.

Specifically, the L.L.L. wavefunction is given by

$$\br \vec x \vert 0,\x ,\a \ket = \sqrt{{{B}\over{2\p }}}
\e^{-{{B}\over 4}\vert 
z\vert^2 + {{B}\over 2}\bar z \x } \eqno(2.22),$$

where $z=x+iy$ and $\bar z$ is its complex conjugate.

The projector onto the L.L.L. is therefore given by

$$ P_0 \equiv {{B}\over {2\pi }}
\dxi \vert 0,\x \ket \br 0,\bar{\x }\vert \eqno(2.23).$$

Thus, from (2.22) and (2.23), 

$$\br \vec x \vert P_0 = \sqrt{{{B}\over {2\p }}}\e^{-{{B}\over 4}\vert z 
\vert^2 }\br 0,\bar z \vert P_0 \eqno(2.24)$$

where $\bar z$, the coherent state parameter in (2.24), is actually constructed
out of the spatial coordinates $x$ and $y$. Namely, $\bar z = x - i y$.

Let us denote by $\vert \y \ket  $ the second-quantised field
operator such that

$$\yh (\vec x,t) \equiv \br \vec x \vert \y (t)\ket  \eqno(2.25).$$

Thus if we project the fields to the L.L.L., from (2.24) and (2.25), we have

$$\yh (\vec x,t)\rightarrow \br \vec x \vert P_0 \vert \y (t)\ket  
\equiv \yh_{0}(\vec x,t)=\e^{-{{B}\over 4}\vert z 
\vert^2 }\yh_{0}(\bar z,t)\eqno(2.26).$$

With this projector $P_0$, we can easily project the entire action to the 
L.L.L.:

$$\eqalignno{
S=\dt \ \dx \ &\bigl[ \yhd_0(\vec x,t)\{ i\del_t -{{\w_c}\over 2}
(1-g S_z )+\m \}\yh_0(\vec x,t)\cr 
&-{1\over 2}\dxx \ \yhd_0(\vec x,t)\yhd_0(\vec x^{\prime },t)
V(\vert \vec x - \vec x^{\prime }\vert )\yh_0(\vec x^{\prime },t)
\yh_0(\vec x,t)\bigr] &(2.27).\cr 
}$$

Using (2.26), the Coulomb term in (2.27) is rewritten as
$$
S_c\equiv -{1\over 2}\dt \zzprexp \yhd_0(z,t)\yhd_0(\zp ,t)V(\ve z-\zp \ve )
\yh_0(\zbarp ,t)\yh_0(\zbar ,t)\eqno (2.28)
$$
where $d^2z \equiv {{B}\over {2\p }}d(\rm Re\ z) d(\rm Im\ z)$.
We should further include a uniform neutralising background charge density
in the system, equal in magnitude to the density associated with a completely
filled L.L.L. Namely, $\r_b = {{B}\over {2\p }}$. Including this explicitly,
the Coulomb term is
$$
S_c = -{1\over 2}\dt \zzprint \bigl[ \zexp \yhd_0(z,t)\yh_0(\zbar ,t)-
{{B}\over {2\pi }}
\bigr] 
V(\ve z - \zp \ve )\bigl[ \zpexp \yhd_0(\zp ,t)\yh_0(\zbarp ,t)-
{{B}\over {2\pi }} \bigr] 
\eqno (2.29).
$$
\bigskip
\magnification=1200
{\bf 3. The Coulomb term and the ground state}
\bigskip
Right at the onset, we should mention that in sections 3 and 4, we shall
omit the suffix 0 from the fermionic fields, as we shall be concerned
only with the L.L.L.

The Coulomb term projected onto the lowest Landau level reads (see appendix B)
$$\eqalign{
L_{c}=& -{1\over 2}\int d^2z_1 d^2z_2
e^{
-{B\over 2}\vert z_1^2\vert
-{B\over 2}\vert z_2^2\vert}
\left( \psi^{\dagger}(z_1)\psi
(\bar z_1)-\rho_0(z_1,\bar z_1) \right) \cr &
V(z_1-z_2, \bar z_1-\bar z_2 )
\left( \psi^{\dagger}(z_2)\psi
(\bar z_2)-\rho_0(z_2,\bar z_2) \right) }
\eqno (3.1)
$$
where the spin indices have been omitted, and the neutralizing
background
charge $\rho_0$ projected onto the lowest Landau level and the Coulomb
potential $V$ are given by
$$\rho_0(z, \bar z)={B\over 2\pi}
e^{
{B\over 2}\vert z^2\vert}
$$
$$
 V(z,\bar z)=
\int {{d^2 k}\over{2\pi }} V(k, \bar k)
e^{i\bar k z + i\bar z k\over 2}
\quad\quad , \quad
V(k, \bar k)={e^2\over \vert
k\vert}
\eqno (3.2)
$$

Notice that the term above is most important at short distances. Let us
then
separate the short distance contribution from the long distance one by
introducing an explicit cut-off $\lambda$, the size of which will be
discussed later on
 $$\eqalign{
 V(z,\bar z)= &
 V_{l}(z,\bar z)+
 V_{s}(z,\bar z) \cr
 V_{l}(z,\bar z)= &
\int^{\lambda}{{d^2 k}\over{2\pi }}  V(k, \bar k)
e^{i\bar k z + i\bar z k\over 2}
\quad\quad , \quad
 V_{s}(z,\bar z)=
\int_{\lambda} {{d^2 k}\over{2\pi }} V(k, \bar k)\cr }
e^{i\bar k z + i\bar z k\over 2}
\eqno (3.3)
$$
Let us assume first that only the short distance contribution is
important
in order to establish the properties of the ground state. We shall prove
later on that this assumption is self-consistent.
We have
$$
L_{c,s}=-{1\over 2}\int d^2z_1 d^2z_2
e^{
-{B\over 2}\vert z_1\vert^2
-{B\over 2}\vert z_2\vert^2}
 \psi^{\dagger}(z_1)\psi
(\bar z_1)
V_{s}(z_1-z_2, \bar z_1-\bar z_2 )
 \psi^{\dagger}(z_2)\psi
(\bar z_2)
\eqno (3.4)
$$
since the neutralising constant charge density gives no contribution for
large (non-zero) momenta. From (3.4) it is not apparent at all that the
spin
degree of freedom is important to establish the nature of the ground
state. Recall that we are interested in the situation where the density
of electrons is equal to the density required to exactly fill up the
lowest Landau level of one of the spin degrees of freedom (which
dictates the choice of the neutralising background charge as given in
(3.2)). However we
cannot guarantee a priori that all the electrons are going to be in the
spin up or down state. As dicussed in the previous section this
situation
is suitably dealt with by the introduction of a density matrix
. 
$$
\rho_{\alpha\beta}(\bar z,z^{\prime})=<\bar z\vert \hat \rho_{\alpha\beta} \vert z^{\prime} >=-
\psi_{\alpha}(\bar z)\psi^{\dagger}_{\beta}(z^{\prime})
\eqno (3.5)
$$
Notice that the mean local density  in nothing but
$$
\br \r (\vec x) \ket = tr(\rho (\bar z,z))e^{{B\over 2}\vert z\vert^2}
\eqno (3.6)
$$
Let us take as an ansatz
$$
\hat \rho_{\alpha\beta}={{B}\over {2\p }}
\int d^2\xi e^{-{B\over 2}\vert \xi\vert^2}
M_{\alpha\beta}(\bar \xi,\xi )
\vert \xi ><\bar \xi\vert
\eqno (3.7)
$$
with $M_{\alpha\beta}(\bar \xi,\xi ) $ being a slowly varying matrix-valued  
function. 
Recall that if 
$M_{\alpha\beta}(\bar \xi,\xi ) $ is constant then the mean local density (3.6) is also
constant. We have, using the results of Appendix A,
$$
\rho_{\alpha\beta}(\bar z,z^{\prime})\sim  e^{-{B\over 2}\bar z z^{\prime}}\left(
M_{\alpha\beta}(\bar z,z^{\prime})+{2\over B}\partial_{z}\partial_{\bar z}M_{\alpha\beta}
(\bar z,z^{\prime} ) +\cdots\right)
\eqno (3.8)  
$$
  
For the Coulomb term (3.4) there are two inequivalent ways of
introducing the density matrix
 $\hat \rho_{\alpha\beta}$ (3.7) 
, namely
$$
\eqalign{
 i)\quad\quad\quad &
 \psi^{\dagger}(z_1)\psi
(\bar z_1)
 \psi^{\dagger}(z_2)\psi
(\bar z_2)
\longrightarrow \rho_{\alpha\alpha}(\bar z_1,z_1)\rho_{\beta\beta}(\bar z_2,z_2) \cr
ii)\quad\quad\quad   &
 \psi^{\dagger}(z_1)\psi
(\bar z_1)
 \psi^{\dagger}(z_2)\psi
(\bar z_2)
\longrightarrow -\rho_{\alpha\beta}(\bar z_1,z_2)\rho_{\beta\alpha}(\bar z_2,z_1) }
\eqno (3.9)
$$
Consider first the case $i)$.
For constant $M_{\alpha\beta}(\bar{\xi} , \xi)$, which
characterises constant
space-time configurations, we obtain 
$$ L_{c,s}= 0                            \eqno (3.10)
$$
This follows immediatly from (3.4) and (3.3). Indeed, upon integrating (3.4) over $z_1$ and 
$\bar z_1$ we obtain a delta function of $k$ and $\bar k$ which, owing to the infrared cut-off
for $V_s(z,\bar z)$ in (3.3), gives zero.
Consider next the case $ii)$. Again, let us choose $M_{\alpha\beta}(\bar z ,z^{\prime})$ 
constant and
take it as
$$ M={\rho_0\over 2}+\vec m\vec \sigma \eqno (3.11)
 $$
We obtain
$$ L_{c,s}={ 2\pi\over B}\int d^2 z
\int_{\lambda} {{d^2 k}\over{2\pi }} V(k, \bar k)
e^{-{\bar k k\over 2B}} \left( {\rho_0^2\over 2}+2{\vec m}^2
\right)
\eqno (3.12)
$$
where the role of the spin is immediate: the system wants to be with
the largest possible constant magnetization $\vec m$
which gives rise to a
negative contribution to the energy. This is only achieved by
having all electrons with the spin pointing in the same direction,
no matter whether the direction is up or down. Namely, we have a
situation of spontaneous symmetry breaking and the ground state is ferromagnetic.
We take, for definiteness,
$$ \vec m=(0,0,{\rho_0\over 2});\  \r_0 \equiv {{B}\over{2\p }} \eqno (3.13)
$$
Since the case $ii)$ leads to a ground state energy lower than the case $i)$ we
understand that 
the choice $ (3.9)\; ii)$ is the
correct one for the short distance piece of the Coulomb interaction. For 
$M_{\alpha\beta}(\bar z ,z^{\prime})$ slowly varying it is easy to 
check that spatial variations
always increase the energy. We shall explicitely display this feature for the magnetic
excitations in the next section.

For the long distance piece of the Coulomb term
 $i)$ still gives rise to a zero contribution to the
energy but $ii)$ leads to ill defined expressions. Therefore for the
long distance part of the Coulomb interaction the ansatz $i)$ should be
used instead of $ii)$.

For the remaining terms in the lagrangian there is only one way to introduce our density matrix
ansatz. Thus for constant configurations we finally obtain

$$L=\mu\rho_0- ({\omega_{c}\over 2}-{gB\over {4m}})\rho_0
+{e^2\rho_0^2\over{2\sqrt{B}}}c 
\eqno (3.14)
$$
$$ c:=4\pi \int_{\lambda\over \sqrt{B}}^{\infty}dk e^{-{k^2\over 2}}  $$ 
Since the Coulomb term gives a contribution to the ground state energy
and we want to have the lowest Landau level of one spin degree of
freedom filled up, we have to tune the chemical potential accordingly.
This is
$$ \mu= ({\omega_{c}\over 2}-{gB\over {4m}})
-{e^2\rho_0 \over{2\sqrt{B}}}c \eqno (3.15)$$


\bigskip

{\bf 4. Effective action for the lowest lying excitations}

\bigskip

Whenever we have a situation of spontaneous symmetry breaking, the
Goldstone theorem ensures that there is no gap in the spectrum [13].
Furthermore, the
lowest lying excitations are quasiparticles (Goldstone bosons) with well
defined transformation properties under the original symmetry group [9].
In
particular they can be described by a slowly varying unitary matrix $U$
living in the
coset $G/H$, where $G$ is the symmetry group of the lagrangian and $H$
that of the ground state. In our case $G=SU(2)$ and $H=U(1)$. Under
$SU(2)$, $U$ transforms non linearly as follows
$$ U(z,\bar z) \rightarrow gU(z, \bar z)h^{-1}(g,U(z,\bar z))  \eqno
(4.1)
 $$
where $g\in SU(2)$ is space-time independent whereas $h\in U(1)$ depends on the space-time
point through $U$.
We use $U$ to built up our ansatz for the lowest lying excitations.
Essentially we specify our density matrix ansatz so that it
includes $U$ and it respects the transformation properties of
$\rho (z,\bar z)$ , namely
$$
\rho (z,\bar z) \rightarrow
g \rho (z,\bar z) g^{-1} \eqno (4.2)
$$
These requirements still leave a certain measure of freedom in the ansatz, 
which is further 
constrained by demanding the
terms with a time derivative in the effective action to be invariant. Up
to and including order $1/B$, this reads
$$ \hat \r =
{B\over 2\pi}
\int d^2\xi e^{-{B\over 2}\bar \xi \xi}
\ddag \tilde U (\hat z ,\hat {\bar z} )\ddag
\vert \xi >\rho_0 p_{+}<\bar \xi\vert
\ddag \tilde U^{\dagger} (\hat z ,\hat {\bar z} )\ddag
$$
$$
\eqalign{
\tilde U(\hat z ,\hat {\bar z} ):= &
 U(\hat z ,\hat {\bar z} )
-{1\over B} \Bigl(\partial_{z}\partial_{\bar z} U(\hat z ,\hat
{\bar z} ) \cr
& -\partial_{z}S\partial_{\bar z} U+\partial_{\bar z}S\partial_{z} U \cr
& +{3\over 2}\partial_{z}S\partial_{\bar z} SU-
{3\over 2}\partial_{\bar z}S\partial_{z} SU \Bigr)
\cr &\cr
& S:=Up_{+}U^{\dagger}
\quad\quad , \quad p_{+}:={1+\sigma_3\over 2}}
 \eqno (4.3)
 $$
Although the time is not explicitly displayed above, $\hat \rho $ may be considered a single
time function for all the terms in the lagrangian except for the time derivative term, where
the time derivative is to be understood as acting on $\tilde U (\hat z ,\hat {\bar z} )$
only and not on ${\tilde U}^{\dagger }(\zh ,\zbh )$.
By using the formulae in Appendix A, one can easily see at this order that
$$\eqalign{
 <\bar z\vert \hat \rho\vert z>=& e^{{B\over 2}\bar z z} \rho_0 \biggl(
Up_{+}U^{\dagger} + \cr  &
+{1\over B}\bigl(\partial_{z}\partial_{\bar z}
(Up_{+}U^{\dagger}) +
(\partial_{z}Up_{+}\partial_{\bar z}U^{\dagger} -
\partial_{\bar z}Up_{+}\partial_{z}U^{\dagger}) \cr
& -2S tr(\partial_{z}Up_{+}\partial_{\bar z}U^{\dagger} -
\partial_{\bar z}Up_{+}\partial_{z}U^{\dagger})\cr
&+\partial_{z}Str(U^{\dagger}\partial_{\bar z}Up_{+})
-\partial_{\bar z}Str(U^{\dagger}\partial_{z}Up_{+})\cr
& -3\partial_{z}S\partial_{\bar z}S S+3\partial_{\bar z}S\partial_{z}S S
 \bigr) \biggr)
}
                                    \eqno (4.4)
$$
which enjoys the desired transformation properties (4.2).
>From (4.4) it follows immediately that the density reads
$$ (\rho (z,\bar z) )_{\alpha \alpha}=\rho_0\left(1+
{2\over B} \Omega \right)
 $$
$$
\Omega:=
tr
(\partial_{z}Up_{+}\partial_{\bar z}U^{\dagger} -
\partial_{\bar z}Up_{+}\partial_{z}U^{\dagger}) 
       \eqno (4.5)
$$
 Upon integration over space, $\Omega$ gives rise to a topological invariant,
the Pontryagin index. Notice that although this term is a total derivative,
it is not a total derivative of an $SU(2)$ invariant object, and hence it should not be
dropped.

Let us next calculate the contributions of the different terms in the
lagrangian to the effective action for $U$.

Consider first the short distance part of the Coulomb term. By
introducing center of mass and relative coordinate we have
$$ \eqalign{
L_{c,s}= & -{1\over 2}\int d^2Z d^2z
e^{
-B\vert Z\vert^2
-{B\over 4}\vert z\vert^2}
 \psi^{\dagger}(Z-{z\over 2})\psi^{\dagger}
( Z+{ z\over 2}) 
V_{s}(z, \bar z)
 \psi(\bar Z+{\bar z\over 2})\psi
(\bar Z - {\bar z\over 2})  \cr
=& \int d^2 Z d^2 z e^{-{B\over 2} \vert z\vert^2} V_{s}(z,\bar z) \cr
& \left( e^{-{z\over 2}\partial_{Z}+{\bar z\over 2}\partial_{\bar Z}}
e^{-{B\over 2}\vert Z\vert^2} \rho_{\alpha\beta}(Z,\bar Z) \right) \cr
& \left( e^{{z\over 2}\partial_{Z}-{\bar z\over 2}\partial_{\bar Z}}
e^{-{B\over 2}\vert Z\vert^2} \rho_{\beta\alpha}(Z,\bar Z) \right) }
\eqno (4.6)
$$
>From the last expression and (4.4) it is apparent that a derivative expansion about the center of
mass coordinates
is well defined. Up to order $1/B$ we obtain
$$
L_{c,s}={1\over 2}  \int d^2Z {e^2\rho_0^2\over \sqrt{B}}c\left( 1+{1\over
B}({b\over c}-2)tr
\left(\partial_{z}S\partial_{\bar z} S \right)+{2\over B}\Omega\right) 
\eqno (4.7).
$$
$c$ has already been defined in (3.14), and
$$
\quad b:=
{B^{3\over 2}\over e^2}\int d^2 z V_{s}(z,\bar z)\vert z\vert^2 
e^{-{B\over 2}\vert z\vert^2}
$$
where $c$ and $b$ are smooth dimensionless functions of $\lambda /\sqrt{B}$ only. Notice that
this term brings up both the kinetic energy for the spin waves and a topological
density. 

The long distance part of the Coulomb interaction gives, by just substituting (4.5) in
(2.29),
$$\eqalign{
L_{c,l}=& \int d^2z_1 d^2z_2
{2\rho_0^2\over B^2}
\Omega (z_1,\bar z_1) V_{l}(z_1-z_2, \bar z_1-\bar z_2 )\Omega (z_2,\bar z_2)
}\eqno (4.8)
$$
which corresponds to a non-local interaction between
topological densities.

The term in the microscopic lagrangian proportional to the density gives (see (2.27))
$$  L_{d}= (-{e^2\rho_0\over 2\sqrt{B}}c -
{gB\over 4m}){2\rho_0\over B}\Omega (z,\bar z)
\eqno (4.9)
$$
where we have substituted (3.15) for the chemical potential. 
Notice that the first term in (4.9) cancels precisely against the last 
term in (4.7). Then, the remaining term that is proportional to the
Pontryagin density is also proportional to the $g$ factor. This 
cancellation was however not observed by the authors of [10], who just
quoted the last term in (4.7) as the only term proportional to the
Pontryagin index.  
The remaining term still  
energetically favours topologically nontrivial configurations, the so
called skyrmions, but the effects are expected to be small on account
of the smallness of $g$. The term proportional to the Pontryagin density
in (4.7) is of the same order in the derivative expansion as the second
term in the same equation. Generically, as we shall see, these two terms
emerge together from the short-distance part of the Coulomb interaction
and topological term has to be retained if the spin-density is. In this
particular instance, the topological term has cancelled out through the
specification of the chemical potential. However, it will be generated
at subleading orders, as will be seen in section 5.
The
topological term, is, of course, odd under parity transformations, but
this is acceptable as the external magnetic field has already broken the
parity symmetry of the system. 


The term with the time derivative gives rise to
$$ L_{t}=\rho_0(A_0-{1\over 2
B}\epsilon^{\mu\nu\rho}A_{\mu}\partial_{\nu}A_{\rho} )$$
$$A_{\mu}:=tr(p_{+} U^{\dagger}i\partial_{\mu} U) 
\eqno (4.10)
$$ 
Notice that both terms are invariant under non-linear $SU(2)$ transformations
 up to total derivatives, and hence are of
topological nature. Moreover, they cannot be
written
in terms of the magnetization in a local form, which shows the obvious
advantage of using $U$ instead of $\vec m$ as the basic field in the
effective
Lagrangian. In fact the second piece in (4.10) is the well known abelian Chern-Simons term
. Its
appearance in QH ferromagnets has been recently pointed out in [10], 
and our result
agrees with that in this article provided we make the identification
$A_{\m }= iz^{\star }\del_{\mu }z$. 
However, the
authors of [10] were not able to write it in terms of the basic fields of the effective action
only, whereas here it appears naturally in this form. Although this term is $1/B$ suppressed
with respect to the first term, requiring invariance of the effective action at this order
constrains very much our ansatz (4.3), which is the main reason why we have displayed it here.
The first term in (4.10) is very interesting \footnote{$^{\dagger}$}{It also arises in
a very
different context: the effective action for the low momentum modes of
heavy quark anti-quark bound states [14], where it was written in the form
(4.10) for the first time.}.
It appears in the literature in a variety of forms, none of them 
being a local expression 
in terms of the
basic fields of the action, as (4.10) is.
 In appendix C we give the
precise relation
between the different forms that this term can be found in.

Finally the Pauli term gives rise to
$$ {gB\over m}{\rho_0\over 8} \left( tr(\sigma_3 U \sigma_3
U^{\dagger})-{2\over B} \Omega - {5\Omega \over{B}}tr(\sigma_3 U \sigma_3
U^{\dagger})
\right)
\eqno (4.11)
 $$
The third term in (4.11) can be regarded as subleading with respect to
the first term and may be dropped. However, the second term, which at
first sight also seems subleading, must be retained owing to the 
cancellation of the leading order term between equations (4.7) and (4.9),
as discussed above. In fact this term combines with the second term in
(4.9) to give a contribution to the Pontryagin density, which however is
proportional to $g$.

Putting all this together we have up to two space and one time derivative
$$
\eqalign{ S_{eff}= &\int dt \Bigl( \int d^2z  \big(  
\rho_0A_0
+
{1\over 2}   {e^2\rho_0^2\over \sqrt{B}}c\left({1\over
B}({b\over c}-2)tr
\left(\partial_{z}S\partial_{\bar z} S \right)\right)
\cr &
-{3 g\r_0\over 4m}\Omega
+
{gB\over m}{\rho_0\over 8} \left( tr(\sigma_3 U \sigma_3
U^{\dagger})-2
\right) \bigr)
\cr &
+\int d^2z_1 d^2z_2
{2\rho_0^2\over B^2}
\Omega (z_1,\bar z_1) V_{l}(z_1-z_2, \bar z_1-\bar z_2 )\Omega (z_2,\bar z_2)
}
\eqno (4.12)
$$

If we take now the cut-off $\lambda$ such that $p << \lambda << \sqrt{B}$ the
coeficients $c$ and $b$ become
$$
c\sim \sqrt{2\pi^3}  \quad\quad  b\sim c 
\eqno (4.13)
$$
and the last term reduces to the Coulomb energy due to the topological
densities. In this situation (4.13) is in agreement with [5], except for 
the topological
density term which remained unnoticed there. 
\bigskip
{\bf 5. Leading quantum corrections due to higher Landau levels}
\bigskip
We show in this section how quantum corrections due to higher Landau
levels can be easily obtained within our formalism. We shall restrict
ourselves to the leading corrections in $\hbar$ ,
$e^2\sqrt{B}/\omega_{c}$
and derivatives.

In appendix B we have obtained the Coulomb term in the Landau level
basis (see (B.7)).
In order to obtain the exact effective action for the spin waves
we should integrate out the fields corresponding to 
all Landau levels but the lowest. This is of
course a formidable task which lies beyond our abilities. Nevertheless, 
we shall be able to
identify the parts of the Coulomb term that are to be retained in
order to obtain the leading corrections we are looking for.

The terms containing three or four fields in higher Landau levels
only
contribute at higher loops ($\ge 2$) and hence they are
supressed by powers of $\hbar$ .
The terms with exactly two fields in higher Landau levels are to be
compared with the terms proportional to the Landau level energies. The
size of the former is at least
$e^2\sqrt{B}/\omega_{c}$
smaller than that of the latter and hence we can also neglect them. 
They would
give rise to corrections to the propagator and to loops. We are then left
with terms with only one field in the higher L.L. These are,
$$
\eqalignno{
L_{c}=& -{1\over 2} \int d^2Zd^2z
\int {{d^2k}\over{2\p }}
{{e^2}\over{\ve k \ve }}\e^{-i{{\bar k}\over 2}z - i {{k}\over 2}\bar z
-B\ve Z \ve^2 -{{B}\over 4}\ve z \ve^2 }
\sum_{n}{1\over{\sqrt{n!}}}\bigl[({k\over{\sqrt{2B}}})^{n}\y_{0}^{\dag }
(Z-{{z}\over 2}) \y_{0}^{\dag }(Z+{{z}\over 2})\cr 
&\times \bigl( \y_{n}(\bar Z + {{\bar z}\over 2}) \y_{0}(\bar Z - {{\bar z}\over 2})
+(-1)^{n}\y_{0}(\bar Z + {{\bar z}\over 2}) \y_{n}(\bar Z - {{\bar z}\over 2})
\bigr) \cr 
&+(-{{\bar k}\over{\sqrt{2B}}})^{n}\bigl( \y_{0}^{\dag }
(\bar Z - {{\bar z}\over 2}) 
\y_{n}^{\dag }(\bar Z + {{\bar z}\over 2})
+(-1)^{n}\y_{n}^{\dag }(\bar Z - {{\bar z}\over 2}) \y_{0}^{\dag }
(\bar Z + {{\bar z}\over 2})
\bigr) \cr 
&\times \y_{0}(\bar Z + {{\bar z}\over 2})\y_{0}(\bar Z - {{\bar z}\over 2})
\bigr] 
& (5.1).\cr 
}
$$

Next, recall that the holomorphic (antiholomorphic) factors must be
compensated for at some point by corresponding antiholomorphic (holomorphic)
factors. Eventually this has to be realised through derivatives on the
the slowly varying fields. Since each derivative goes with a factor
$1/\sqrt{B}$, the more important terms are those with a smaller number 
of unpaired
holomorphic (antiholomorphic) factors. In other words, the first Landau
level fields are the most important ones, which is very reasonable from
a physical point of view. The $n$th Landau level gives a
contribution suppressed by $(p/\sqrt{B})^{n}$, where $p$ is the typical
momentum of the spin waves. Then the sum over $n$ in (5.1) reduces to $n=1$.
 
Next we separate, as we did in the previous section, the
Coulomb term into short distance and long distance contributions, and
organise the fields accordingly. For the short dictance part we have
$$\eqalign{
L_{c,s}= & {1\over 2}\int d^2Z d^2z
\int_{\lambda } {d^2 k \over 2\pi}{e^2\over \vert k\vert }
e^{-i{{\bar k}\over 2}z - 
i{{k}\over 2}\bar z }
e^{
-{B\over 2}\vert Z\vert^2
-{3B\over 8}\vert z\vert^2}
\cr &
tr\Big(
-e^{B{z\bar Z\over 4}-B{\bar z Z\over 4}}
 \left( e^{{z\over 2}\partial_{Z}-{\bar z\over 2}\partial_{\bar Z}}
e^{-{B\over 2}\vert Z\vert^2} \rho (Z,\bar Z) \right)
\cr &
\left({ k\over \sqrt{2B}} \psi_{1}(\bar Z+{\bar z\over 2})
 \psi^{\dagger}_{0}(Z-{z\over 2}) 
+{\bar k\over \sqrt{2B}}
\psi_{0}(\bar Z+{\bar z\over 2})
 \psi^{\dagger}_{1}
( Z-{ z\over 2}) 
 \right) 
\cr &  +
e^{-B{z\bar Z\over 4}+B{\bar z Z\over 4}}
\left( e^{-{z\over 2}\partial_{Z}+{\bar z\over 2}\partial_{\bar Z}}
e^{-{B\over 2}\vert Z\vert^2} \rho (Z,\bar Z) \right)
\cr &
\left(
{ k\over \sqrt{2B}} \psi_{1}
(\bar Z - {\bar z\over 2})
 \psi^{\dagger}_{0}(Z+{z\over 2}) 
+{\bar k\over \sqrt{2B}}
\psi_{0}
(\bar Z - {\bar z\over 2})
 \psi^{\dagger}_{1}
( Z+{ z\over 2}) 
  \right)  \Big)
\cr &\cr 
\sim &
{ie^2 c\over 4\pi \sqrt{2}}
\int d^2 Z e^{-{B\over 2}\vert Z\vert^2}
\left(
 \psi^{\dagger}_{0}(Z)\partial_{\bar Z}S 
 \psi_{1}(\bar Z)
-\psi^{\dagger}_{1}(Z)\partial_{ Z}S 
 \psi_{0}(\bar Z) \right)}
\eqno (5.2)
$$
and for the long distance part
$$\eqalign{L_{c,l}=&
-
\int d^2z_1 d^2z_2
e^{
-{B\over 2}\vert z_1\vert^2
-{B\over 2}\vert z_2\vert^2}
\int^{\lambda} {d^2 k \over 2\pi}
{e^2\over \vert k\vert }e^{-i{{\bar k}\over 2}(z_1 - z_2)
-i{{k}\over 2}(\bar z_1-\bar z_2 )}
\cr &
\left( \psi^{\dagger}_0(z_2)\psi_0
(\bar z_2)-
{{B}\over{2\p }}e^{
{B\over 2}\vert z^2_2\vert}\right)
\left(
 {k\over \sqrt{2B}}
 \psi^{\dagger}_{0}(z_1)
 \psi_{1}
(\bar z_1 ) 
-{\bar k\over \sqrt{2B}}
 \psi^{\dagger}_{1}
(  z_1) 
 \psi_{0}
(\bar z_1)\right) 
\cr &\cr
\sim & -{1\over \pi }
\int d^2z_1 d^2z_2
e^{
-{B\over 2}\vert z_1\vert^2
-{B\over 2}\vert z_2\vert^2}
\int^{\lambda} {d^2 k \over 2\pi}{e^2\over \vert k\vert }e^{-i{{\bar k}\over 2}(z_1 - z_2)
-i{{k}\over 2}(\bar z_1-\bar z_2) }
\cr &
\Omega(z_2,\bar z_2)\left(
 {k\over \sqrt{2B}}
 \psi^{\dagger}_{0}(z_1)
 \psi_{1}
(\bar z_1 ) 
-{\bar k\over \sqrt{2B}}
 \psi^{\dagger}_{1}
(  z_1) 
 \psi_{0}
(\bar z_1)\right) }
\eqno (5.3)
$$
where we have reinserted the background neutralising charge. Here again the background
charge does not contribute to the  short distance part but it is 
essential for obtaining a well
defined long distance part.

The terms in the lagrangian which are quadratic in the first Landau level field read
$$
\int d^2z
e^{
-{B\over 2}\vert z\vert^2}
 \psi^{\dagger}_{1}
(  z) (i\partial_0-\omega_{c})
 \psi_{1}
(\bar z)
\eqno (5.4)
$$
where we have dropped subleading contributions due to the Pauli term. The functional
integral over the first Landau level field is now gaussian and can be readily carried out.
This leads to the following extra term in the effective action
$$
\eqalign{L_{eff}^{1}= &
\int d^2z 
e^{
-{B\over 2}\vert z\vert^2}
\cr &
\Bigl(
 -{2\over \pi}
\int d^2z_2
\int^{\lambda} {d^2 k \over 2\pi}{e^2\over \vert k\vert }e^{-i{{\bar k}\over 2}(z_2 - z)-
i{{k}\over 2}(\bar z_2-\bar z) }
\Omega(z_2,\bar z_2)
 {k\over \sqrt{2B}}
 \psi^{\dagger}_{0}(z)  
\cr &
-i{ e^2 c\over 4\pi \sqrt{2}}
\psi^{\dagger}_{0}(z)\partial_{\bar z}S 
 \Bigr)\left( -{1\over i\partial_0-\omega_{c} }\right)
\Bigl(
i{ e^2 c\over 4\pi \sqrt{2}}
\partial_{ z}S 
 \psi_{0}(\bar z)
\cr &
 +{2\over \pi }
\int d^2z_2
\int^{\lambda} {d^2 k \over 2\pi}{e^2\over \vert k\vert }e^{-i{{\bar k}\over 2}(z_2 - z)-
i{{k}\over 2}(\bar z_2-\bar z )}
\Omega(z_2,\bar z_2)
{\bar k\over \sqrt{2B}}
 \psi_{0}
(\bar z)\Bigr)}
\eqno (5.5)
$$
Since we consider spin configurations slowly varying in time,
we can neglect the time derivatives in the denominators. We then obtain at leading order 
in derivatives
$$
L_{eff}^{1}\sim -{Be^4c^2\over {64 \pi^3 \omega_{c} }}
tr(S\partial_{\bar z}S\partial_{z}S)
\eqno (5.6)
$$
This term, which arises from the short distance contributions,
amounts to a finite renormalization of the kinetic and topological term
respectively. The leading term arising from the long distance
contribution, becomes local if we take $\lambda$ much larger than
$p$, the typical momentum of the spin waves. It reads
$$
{e^4\over \pi\omega_{c}}\int d^2 z\  \Omega^2(z,\bar z)
\eqno (5.7)
$$
This term, however, is
subleading
with respect to four-derivative terms that arise from the lowest
Landau level alone, when we improve our ansatz further. The
cross terms in (5.5) involving short and long distance contributions vanish.
Let us finally mention that a subleading terms in the long distance
contribution of (5.5) gives rise to a surprising non-local three-body interaction between
the topological densities which reads
$$
{1\over \pi^3 \omega_{c}} \int d^2z_1 d^2z_2 d^2 z
\partial_{ z_2}\Omega(z_2,\bar z_2)V(z_2-z,\bar z_2 -\bar z)\Omega(z,\bar z)
V(z-z_1,\bar z -\bar z_1)
\partial_{\bar z_1}\Omega(z_1,\bar z_1)
\eqno (5.8)
$$

The bottom line of this section is that
higher Landau levels produce no qualitative
change to the results obtained upon projection to the L.L.L. : only a finite renormalization of the kinetic energy and the
topological density terms is encountered at the lowest order in derivatives. 
At higher orders, though, qualitatively new terms like
(5.8) arise. It is important to remark that for spin waves with very low momentum $p$ or for very large skyrmions,
 the corrections due
to higher Landau levels that we have just calculated, are the dominant 
ones. Namely, they go as $e^2\sqrt{B}/\omega_{c}$ against 
$p/\sqrt{B}$, which is the order of the corrections to our ansatz. 
A calculation of the latter may be found in [10].
\bigskip
\magnification=1200
\overfullrule=0pt
\baselineskip=20pt
\parskip=0pt
\def\dag{\dagger}
\def\del{\partial}

\def\a{\alpha}             
\def\b{\beta}              
        
\def\d{\delta}        
\def\e{{\rm e}}           
\def\z{\zeta}              
\def\j{\eta}

\def\m{\mu}	           
\def\n{\nu}                
\def\x{\xi}              
                  
\def\p{\pi}        \def\P{\Pi}      
\def\r{\rho}

\def\y{\psi}

\def\yh{\hat{\y}}
\def\yhd{\hat{\y}^{\dag}}

\def\w{\omega}     
   
\def\br{\langle}

\def\ve{\vert}

\def\zbar{\bar{z}}

\def\eqref#1{(\ref{#1})}

\magnification=1200
{\bf 6. Discussions and Conclusions}
\bigskip
In this article, we have addressed the issues of spontaneous symmetry
breaking in integral quantum Hall systems, the emergence of magnons as
the corresponding Goldstone modes and the existence of charged, topologically
nontrivial spin configurations (spin skyrmions) as the lowest-lying charge
carriers in the system, when the filling fraction deviates marginally from 
$\nu = 1$. 

The starting point has been the microscopic non-relativistic fermionic action
that governs a system of planar electrons in a magnetic field orthogonal to
the plane. The Coulomb interaction, which normally plays a subsidiary role
in discussions of the integral Hall effect, is included from the beginning,
as it is expected that the experimentally measured gap in the spectrum is
generated precisely by this term in the case of odd-integer fillings.

Contrary to most other approaches to this problem, we have adopted a second-
quantised framework in this article, as S.S.B. is handled with the greatest
facility with the machinery of quantum field theory. Consequently, we have
avoided overt references to many-body wave functions. The information normally
carried by these has been supplied through the specification of the density
matrix, describing the ground state, and further on, the lowest excited
states of the system. Once the density matrix has been provided, the computation
of the physical properties of the system is rendered rather tractable. In our
case, we have made a very natural ansatz for the density matrix describing
the $\nu = 1$ many-body ground state. The mean energy of the ground state is
then computed and we discover that this energy is minimised if the ansatz
is adjusted to describe electrons in the L.L.L., with their spins
\underbar{{\bf \it spontaneously aligned in a given direction,}}
even when the Land\'e g-factor is zero. This is thus an unambiguous
demonstration of S.S.B. In fact, within our approach, it is easy to see that
it is the Coulomb interaction, projected onto the L.L.L. which is responsible
for this spontaneous alignment of the spins. Thus, as a first step, we have
established that the ground state of the $\nu = 1$ system is ferromagnetic.
This result is quite well-known. However, we believe that our approach possesses
the merits of directness and clarity.

Having established the nature of the ground state, we have generalised the
ansatz to include the lowest-lying excitations, both neutral and charged.
This is done, using a technique quite familiar from high energy physics.
The density matrix of the excited state is the unitary transform of that
describing the ground state, where the matrices implementing this transformation
live in the coset space $G/H$, where $G$ is the original symmetry group
and $H$ the unbroken subgroup. The density
matrix has to be covariant under the original symmetry group. In our case,
since the projection onto the L.L.L. naturally entails the non-commutativity
of the components of the coordinate vectors, the unitary transformation has
been carefully defined to ensure this covariance, to a given order in the
natural expansion parameter, ${1\over{B}}$. 

What this unitary transformation actually describes is the local rotation of
the spins from their fixed direction in the ferromagnetic ground state.
Hence, the mean microscopic action, computed with this new density matrix,
naturally yields the effective action governing the spin dynamics. 

Since the ground state is ferromagnetic, the magnons should, as is well
known, possess a quadratic dispersion relation. This in turn means that
the time derivative term in the effective Lagrangian should be linear.
In fact, we obtain the  term with one time
derivative in a simple and covariant form, which is equivalent to more
conventional forms found in the literature, as we have proved in 
appendix C. In this appendix, we also show that this term can be cast 
into the form of a two dimensional Wess-Zumino term. 
As we have
mentioned, the spin alignment in the ferromagnetic ground state is
wrought by the short-distance part of the Coulomb interaction. This
correspondingly, would imply that the short-distance part of the
Coulomb interaction should yield the spatial gradient term in the
effective spin Lagrangian, which measures the energy required to
bring the spins out of alignment. This is indeed what emerges from
our calculations.

As we have noted, the projection to the L.L.L. leads
to a very interesting feature, namely, a local alteration of the
charge density is automatically accompanied by a local change in the
spin density. Turning the argument around, if one creates a topologically
nontrivial spin texture, this configuration would carry charge. We show,
very simply and directly, that our ansatz for the excited state leads 
immediately to a charge density that could yield a nontrivial topological
charge with a non-zero Pontryagin index. The O(3) non-linear $\sigma $
model that describes the magnons is well-known to possess skyrmionic
solutions. What is novel in the Hall systems is that these skyrmions
are charged. They have an electric charge which equals their topological
charge. Moreover, in the effective Lagrangian, we find a term proportional
to the topological density, which energetically favours topologically
non-trivial configurations. However, this term is also proportional to
the $g$ factor and hence is expected to give a small contribution. 
We obtain an interaction among these charged topological
objects mediated by the long-distance part of the Coulomb interaction.
It is also worth pointing out that the C-S term that we have obtained
depends in a simple way on the basic field $U$, so that no auxiliary
variable is needed, unlike in [10], to build it up.

Most of these above features have been noted in other works, but not all
together within a single framework. In [3,5], the time derivative, kinetic
and long-range interaction terms were found, but the topological density 
term was missing, as well as the Hopf term. On the other hand, our 
results essentially agree with those of [10], except for the coefficient
of the Pontryagin density. This is because a crucial cancellation had
remained unnoticed by these authors.  
In fact, in order to
obtain all the terms, it is important to separate the Coulomb interaction 
between high and low momentum pieces. 
An important feature of our approach, which we have focussed
on in the previous section, is the facility it affords in including the
effects of the higher L.L., which have been totally ignored in previous
works. In principle, our second-quantised approach incorporates a
systematic expansion in the various parameters $\sqrt{B}, \w_{c} , 
e^2\sqrt{B}$. For low energy spin waves, say with energy $E$, or for
skyrmions with a large size, these corrections, $\sim {{e^2\sqrt{B}}
\over{\w_{c} }}$, are more important than those coming from higher order
terms in the derivative expansion, $\sim {{E}\over {e^2\sqrt{B}}}$.
This means that
we were able to compute the effects of higher L.L. mixing due to the
Coulomb term, rather straightforwardly. These lead to a
renormalisation of the parameters of the leading order effective
Lagrangian. This is not easily accomplished by the other methods
currently available in the literature. This aspect becomes even more
important when the electrons are coupled to electromagnetic probes.
After all, the skyrmions, as charged objects are touted as the 
favoured charge carriers in the $\nu = 1$ system and one would like
to investigate their electromagnetic responses. The electromagnetic
fields would however set up virtual transitions and consequent mixing
of the Landau levels. We thus expect our method, which is geared
towards including such mixings, to be invaluable in this situation.
Work in this direction is currently in progress.

In conclusion, we may summarise the contents of this article as follows.
We have used field theoretic techniques to demonstrate that the ground
state of the $\nu = 1$ system is ferromagnetic. Furthermore, we have
extracted the effective action governing the resultant Goldstone modes,
the magnons. In this effective action, we have found several topological
terms. The leading term with a single time derivative, the Chern-Simons
(Hopf) term and the term proportional to the Pontryagin index.  
Moreover, we have established
a systematic algorithm, whereby contributions to the spin effective
action, which arise from L.L. mixing, may be systematically computed.
This, we reiterate, is the most significant part of this work.   
\bigskip
\centerline {\bf Acknowledgements}
\bigskip
We have benefitted from informative discussions with L. Brey, B. Sakita
,R. Cot\' e and H. Hansson. J.S. thanks W. Apel for discussing his work prior to
publication. The authors would like to thank the organisers of the 
Benasque Centre of Physics, for providing a nice atmosphere in which
this work could be written up. Financial support from the NSERC and 
from Grant Nos. CICYT AEN95-0590 and GRQ93-1047 is acknowledged.
\bigskip
\magnification=1200
\overfullrule=0pt
\baselineskip=20pt
\parskip=0pt
\def\dag{\dagger}
\def\del{\partial}

\def\a{\alpha}             
\def\b{\beta}              
        
\def\d{\delta}        
\def\e{{\rm e}}           
\def\z{\zeta}              
\def\j{\eta}

\def\m{\mu}	           
\def\n{\nu}                
\def\x{\xi}              
                  
\def\p{\pi}        \def\P{\Pi}      
\def\r{\rho}

\def\y{\psi}

\def\yh{\hat{\y}}
\def\yhd{\hat{\y}^{\dag}}

\def\w{\omega}     
   
\def\br{\langle}

\def\ve{\vert}

\def\zbar{\bar{z}}

\def\eqref#1{(\ref{#1})}

\magnification = 1200
\centerline{\bf References}
\bigskip
\bigskip
\item{[1]} B.I. Halperin, Helv. Phys. Acta {\bf 56}, 75, (1983)
\item{[2]} A.H. MacDonald, cond-mat/9410047
\item{[3]} S.L. Sondhi, A. Karlhede and S.A. Kivelson, Phys. Rev. {\bf B 47},
16419, (1993)
\item{[4]} S.E. Barrett et al, Phys. Rev. Lett. {\bf 74}, 5112, (1995)
\item{[5]} K. Moon et al, Phys. Rev. {\bf B 51}, 5138, (1995)
\item{[6]} Michael Stone, cond-mat/9512010
\item{[7]} T. Hansson, A. Karlhede and J.M. Leinaas, cond-mat/9606116
\item{[8]} H.A. Fertig et al, cond-mat/9612210
\item{[9]} S. Coleman, J. Wess and B. Zumino, Phys. Rev. {\bf 177}, 2239, 
(1969)
\item{[10]} W. Apel and Yu. A. Bychkov, cond-mat/9610040
\item{[11]} F. Wilczek and A. Zee, Phys. Rev. Lett. {\bf 51}, 2250, (1983)
\item{[12]} R. Ray and J. Soto, Phys. Rev. {\bf B 54}, 10709, (1996)
\item{[13]} J. Goldstone, A. Salam and S. Weinberg, Phys. Rev. {\bf 127},
965, (1962); G.S. Guralnik, C.R. Hagen and T.W.B. Kibble, {\it Advances
in Particle Physics}, Vol. {\bf 2}, 567, (Wiley, New York), (1968)
\item{[14]} A. Pineda and J. Soto, Phys. Rev. {\bf D 53}, 3983, (1996)
\item{[15]} H. Leutwyler, Phys. Rev. {\bf D 49}, 3033, (1994)
\vfill
\eject

\magnification=1200
\overfullrule=0pt
\baselineskip=20pt
\parskip=0pt
\def\dag{\dagger}
\def\del{\partial}

\def\a{\alpha}             
\def\b{\beta}              
        
\def\d{\delta}        
\def\e{{\rm e}}           
\def\z{\zeta}              
\def\j{\eta}

\def\m{\mu}	           
\def\n{\nu}                
\def\x{\xi}              
                  
\def\p{\pi}        \def\P{\Pi}      
\def\r{\rho}

\def\y{\psi}

\def\yh{\hat{\y}}
\def\yhd{\hat{\y}^{\dag}}

\def\w{\omega}     
   
\def\br{\langle}
\def\ket{\rangle}
\def\ve{\vert}

\def\zbar{\bar{z}}

\def\eqref#1{(\ref{#1})}
\def\zexp{\e^{-{{B}\over 2}\ve z \ve^2 }}

\def\enorm{\e^{{2\over{B}}\del_{\eta }\del_{\bar{\z }}}}

\def\zpexp{\e^{-{{B}\over 2}\ve \zp \ve^2 }}

\def\zp{z^{\prime }}
\def\zbarp{\bar{\zp }}
\def\zzprint{\int d^2z d^2\zp }
\def\zzprexp{\zzprint \e^{-{{B}\over 2}(\vert z \vert^2 +\vert \zp \vert^2 )}}

\def\dt{\int dt}
\def\dx{\int d\vec x}
\def\dxx{\int d\vec x^{\prime } }
\def\dxi{\int d^2 \x \ \e^{-{{B}\over 2}{\vert \x \vert^2}}}

\def\Phx{\hat {\P^x }}
\def\Phy{\hat {\P^y }}

\def\Ph{\hat \P }
\def\Phd{\hat {\P^{\dag }}}
\def\Xh{\hat X}
\def\Yh{\hat Y}
\def\zh{\hat z}
\def\zbh{\hat{\zbar }}
\def\Zh{\hat Z}
\def\Zbh{\hat {\bar Z}}

\magnification=1200
{\bf  Appendix A}
\bigskip
In this appendix, we shall elaborate upon some results used 
extensively in the main text.
\bigskip
{\bf a.}
\bigskip
$$
\br \bar \z \ve \sharp f(\Zh ,\Zbh )\sharp \ve \eta \ket = 
\enorm \br \bar \z \ve : 
f(\Zh ,\Zbh ): \ve \eta \ket \eqno (A.1).
$$
This is shown as follows:
$$
f(\Zh ,\Zbh )=\int d\vec k f(\vec k)\e^{i{{\bar k}\over 2}\Zh + 
i{{k}\over 2}\Zbh } \eqno (A.2).
$$
Thus, by definition,
$$
\sharp f(\Zh ,\Zbh )\sharp = \int d\vec k \e^{i{{\bar k}\over 2}\Zh }
\e^{i{{k}\over 2}\Zbh }f(\vec k) \eqno (A.3).
$$
Using the result
$$
\e^{A} \e^{B} = \e^{B} \e^{A} \e^{[A,B]}
$$
with
$A\equiv i{{\bar k}\over 2}\Zh $, $B\equiv i{{k}\over 2}\Zbh $ and 
$[A,B]=-{{{\vec k}^2}\over {2B}}$, we get, from (A.3),
$$
\sharp f(\Zh ,\Zbh ) \sharp =\int d\vec k \e^{-{{{\vec k}^2}\over {2B}}}
\e^{i{{k}\over 2}\Zbh } \e^{i{{\bar k}\over 2}\Zh }f(\vec k)  
\eqno (A.4). 
$$
Thus, sandwiching (A.4) between coherent states, we get (A.1) directly.
\bigskip
{\bf b.}
\bigskip
Another expression that we repeatedly encounter is:
$$
A_{f,g}\equiv \br \bar z \ve \bigl[ \dxi : f(\Zh ,\Zbh ) : \ve 0,\xi \ket 
\br 0,\bar \xi \ve : g(\Zh ,\Zbh ) : \bigr] \ve z \ket \eqno (A.5).
$$
Since we know the action of $\Zh $ and $\Zbh $ on the coherent states,
we obtain from (A.5)
$$
A_{f,g}=\int d^2\xi \e^{-{{B}\over 2}\ve \xi \ve^2 + {{B}\over 2}\bar z
\xi + {{B}\over 2}\bar \xi z}f(\xi ,\bar z)g(z,\bar \xi ) \eqno (A.6).
$$
Shifting, $\xi = \j + z$ and $\bar \xi = \bar \j + \bar z$, we get
$$
A_{f,g}=\e^{{{B}\over 2}\ve z \ve^2 }\int d^2 \j \e^{-{{B}\over 2}
\ve \j \ve^2 }f(z+\j ,\bar z)g(z,\bar z + \bar \j ) \eqno (A.7),
$$
which may be expanded as
$$
A_{f,g}=\e^{{{B}\over 2}\ve z \ve^2 }[f(z,\bar z)g(z,\bar z)
+{2\over {B}}\del_z f\del_{\bar z}g + \cdots ] \eqno (A.8),
$$
where the dots indicate terms of $O({1\over {B^2}})$.
\bigskip
{\bf Appendix B}
\bigskip
In section 5, we have described the contributions to the effective
action for the fermionic spin, arising from the mixing of the
higher L.L. with the L.L.L. This mixing is brought about by the 
Coulomb term. In this appendix, we derive an expression for the
Coulomb term, projected onto the L.L. basis.

The bare Coulomb term may be expressed as 
$$
V(\hat{\vec r_1 } - \hat{\vec r_2 }) = \int {{d\vec k }\over {2\pi }}
\e^{i\vec k \cdot (\hat{\vec r_1 } - \hat{\vec r_2 })}{{e^2}\over{\ve \vec k \ve 
}}
\eqno (B.1)
$$
where $\vec r_{i} \equiv (x_{i} , y_{i} ).$

Thus, the Coulomb Lagrangian is 
$$\eqalignno{
L_c = -\int {{d\vec k }\over {4\pi }}{{e^2}\over{\ve \vec k \ve }}
\sum_{n_1,n_2,n_1^{\prime },n_2^{\prime }=0}^{\infty }
&\int \prod_{i=1}^{2} d^2\x_{i} d^2\x_{i}^{\prime }
\e^{-{{B}\over 2} (\ve \x_{i} \ve^2 + \ve \x_{i}^{\prime }
\ve^2 )}W_{n_1,n_1^{\prime },n_2,n_2^{\prime }}(\vec k) \cr 
&\yhd_{n_1}(\x_1)\yhd_{n_2}(\x_2)\yh_{n_2^{\prime }}
(\bar{\x_2^{\prime }})
\yh_{n_1^{\prime }}(\bar{\x_1^{\prime }}) & (B.2)\cr 
}
$$
where 
$$
W_{n_1,n_1^{\prime },n_2,n_2^{\prime }}(\vec k) 
\equiv \br n_1,\bar{\x }_1 \ve \br n_2,\bar {\x }_2 \ve 
\e^{-i\vec k \cdot (\hat{\vec r_1}-\hat{\vec r_2})}\ve n_2^{\prime },
\x_2^{\prime }\ket \ve n_1^{\prime },\x_1^{\prime }\ket 
\eqno (B.3).
$$
Expressing the coordinate operators in terms of $\Zh , \Zbh , \Ph , \Phd $
we get
$$
W_{n_1,n_1^{\prime },n_2,n_2^{\prime }}(\vec k) = 
\e^{-i{{k}\over 2}(\bar{\x_1}-\bar{\x_2})-i{{\bar{k}}\over 2}
(\x_1^{\prime }-\x_2^{\prime })}\e^{{{B}\over 2}(\bar{\x_1}\x_1^{\prime }+
\bar{\x_2}\x_2^{\prime })}M_{n_1,n_1^{\prime }}(\vec k)M_{n_2,n_2^{\prime }}
(-\vec k)
\eqno (B.4).
$$
Here,
$$
M_{n,n^{\prime }}(\vec k)\equiv \br n \ve \e^{{{k}\over{2B}}\Ph }
\e^{-{{\bar{k}}\over{2B}}\Phd }\ve n^{\prime }\ket 
\eqno (B.5),
$$
with $k\equiv k_{x}+ik_{y}$ and $\bar{k}\equiv k_{x}-ik_{y}$.

Upon simplification,
$$
M_{n,n^{\prime }}(\vec k)={1\over{\sqrt{n!n^{\prime }!}}}
(-{{\bar{k}}\over{\sqrt{2B}}})^{n-n^{\prime }}
\sum_{l=max(0,n^{\prime }-n)}^{\infty }{{(n+l)!}\over{l!(n-n^{\prime }+l)!}}
(-{{k^2}\over{2B}})^{l}
\eqno (B.6).
$$
Using $\e^{{{B}\over 2}(\bar{\x }_2 + {{i\bar k}\over {B}} )\x_2^{\prime }}
= \br \bar{\x }_2 + {{i\bar k}\over {B}}\ve \x_2^{\prime } \ket  $,
we get,
$$
\e^{-{{B}\over 2}\ve \x_2 \ve^2 + i{{\bar k}\over 2}\bar{\x }_2 }
\int d^2\x_2^{\prime }\ \e^{-{{B}\over 2}\ve \x_2^{\prime } \ve^2 }
\e^{{{B}\over 2}(\bar{\x }_2+{{i\bar k}\over{B}})\x_2^{\prime }}
\yh_{n_2^{\prime }}(\bar {\x }_2^{\prime })
= \e^{{{k^2}\over {2B}}-{{B}\over 2}\ve \x_2 \ve^2 + i{{\bar k}\over 2}\x_2 
+ i{{k}\over 2}\bar {\x }_2}\yh_{n_2^{\prime }}(\bar {\x }_2) .
$$
Using this to integrate the primed variables out, we obtain, from (B.2),
$$\eqalignno{ 
L_c = -e^2 \int {{d\vec k }\over {4\pi }}{{\e^{{{k^2}\over{B}}}}
\over{\ve \vec k \ve }}
&\sum_{n_1,n_2,n_1^{\prime },n_2^{\prime }=0}^{\infty }
M_{n_1,n_1^{\prime }}(\vec k)M_{n_2,n_2^{\prime }}(-\vec k)
\int \prod_{i=1}^{2}d^2\x_{i}\e^{-{{B}\over 2}\ve \x_{i} \ve^2 }
\e^{-i{{k}\over 2}(\bar{\x_1}-\bar{\x_2})-i{{\bar{k}}\over 2}
(\x_1-\x_2 )}\cr 
& \yhd_{n_1}(\x_1)\yhd_{n_2}(\x_2)\yh_{n_2^{\prime }}
(\bar{\x_2})\yh_{n_1^{\prime }}(\bar{\x_1})
& (B.7).\cr 
}$$
This is the Coulomb Lagrangian, projected onto the L.L. basis. The
term with all the L.L. indices set to zero is the Coulomb interaction
projected onto the L.L.L. and has been used extensively in the 
earlier sections of the article. 

>From (B.6), it is easily seen that,
$$
M_{n,n}(\vec k)=M_{n,n}(-\vec k)={1\over{n!}}{{\del^n }
\over{\del \r^{n}}}\ve_{\r =-{{k^2}\over {2B}}}(\r^{n}\e^{\r })
=\e^{-{{k^2}\over{2B}}}+ O(1/B)
\eqno (B.8)
$$
for all $n\ge 0$.
Again, for $n\neq 0$, we have
$$
M_{0,n}(\vec k)={1\over{\sqrt{n!}}}({{k}\over{\sqrt{2B}}})^{n}
\e^{-{{k^2}\over{2B}}};\ 
M_{n,0}(\vec k)={1\over{\sqrt{n!}}}({{-\bar{k}}\over{\sqrt{2B}}})^{n}
\e^{-{{k^2}\over{2B}}}
\eqno (B.9).
$$
This shows quite clearly that mixing between different Landau levels
is suppressed by powers of ${1\over{\sqrt{B}}}$ and hence, if we
look at the mixing of higher L.L. with the L.L.L., it would suffice
to consider only $n=1$.

\bigskip
{\bf Appendix C}
\bigskip

In this appendix we make contact with various representations of the
Goldstone fields found in the literature.

Let us start with $M=U\sigma^3 U^{\dagger}$ and recall that
$$
S=Up_{+} U^{\dagger}={1\over 2}+{1\over 2}M \quad\quad ,\quad
 M^{\dagger}=M
 \quad\quad ,\quad
 trM=0 \quad\quad , \quad M^2=1
 \eqno (C.1)
$$
Then we can write
$$
M= \vec \sigma \vec m
 \quad\quad ,\quad
{\vec m}^2=1
\eqno (C.2)
$$
Since $M\rightarrow g M g^{-1}$ , $\vec m$ transforms in the vector
representation of $SU(2)$. In terms of $\vec m$, for instance, the
kinetic energy term reads
$$
4tr(\partial_{z} S\partial_{\bar z} S)=
tr(\partial_{z} M\partial_{\bar z} M)=2
\partial_{z} \vec m\partial_{\bar z} \vec m
\eqno (C.3)
$$
It is also very easy to check that
$$
2\Omega =
tr
(\partial_{z}U\sigma^3\partial_{\bar z}U^{\dagger} -
\partial_{\bar z}U\sigma^3\partial_{z}U^{\dagger})
=-{i\over 8}
\epsilon^{\alpha\beta}
tr(M
\partial_{\alpha}M
\partial_{\beta}M)
={1\over 4}
\epsilon^{\alpha\beta}
\epsilon_{ijk} m^{i}
\partial_{\alpha}m^{j}
\partial_{\beta}m^{k}
\eqno (C.4)
$$
$(\alpha , \beta = x , y )$
 which is a more conventional form of the
topological density.

Next, consider the term with the time derivative
$$
T=2A_0 =
tr(\sigma_3 U^{\dagger}i\partial_0 U)
\eqno (C.5)
$$
As mentioned before this term cannot be written in terms of $S$ or $\vec
m$ in a local form. However, we can write it locally in $S$ or $\vec m$
if we introduce an extra dimension in the following way. We interpolate
smoothly the Goldstone fields
$\pi^{\alpha}(x)\rightarrow
\pi^{\alpha}(x,\lambda)
\quad , \lambda\in [0,1]$ in such a way that
$
\pi^{\alpha}(x,1)=
\pi^{\alpha}(x)$
and
$\pi^{\alpha}(x,0)=0 $
. Then it is very easy to check that
$$
T=-{i\over 4}
\int_0^1d\lambda
\epsilon^{\alpha\beta}
tr(M
\partial_{\alpha}M
\partial_{\beta}M)
=
{1\over 2}
\int_0^1d\lambda
\epsilon^{\alpha\beta}
\epsilon_{ijk} m^{i}
\partial_{\alpha}m^{j}
\partial_{\beta}m^{k}
\eqno (C.6)
$$
$(\alpha , \beta = 0 , \lambda )$.
To our knowledge, the last expresion for $T$ was first written down in [15], whereas 
both the second last expression, which resembles a two dimensional Wess-Zumino term, and
(C.5) are original.
  $T$ is usually found with a rather different aspect, namely,
$$
T=\vec A (\vec m)\partial_0 \vec m
 \quad\quad ,\quad
\epsilon_{ikj}
{\partial A^{k}\over \partial m^{j}}=
m^{i}
\eqno (C.7)
$$
It is simple to check that (C.7) and (C.6) lead to the same equation of motion.

\end



Finally, if we want to write the lagrangian in terms of properly
normalized fields, we just have to expand
$$
U=e^{i\pi_{\alpha}\sigma^{\alpha}\over f_{\pi}}=1+
i{\pi_{\alpha}\sigma^{\alpha}\over f_{\pi}}+ \cdots
 \quad\quad ,\quad
\alpha=1,2
\eqno (C.9)
$$
 The lagrangian to quadratic order reads
$$
  \eqno (C.10)$$
^Z

\magnification=1200
\overfullrule=0pt
\baselineskip=20pt
\parskip=0pt
\def\dag{\dagger}
\def\del{\partial}

\def\a{\alpha}             
\def\b{\beta}              
        
\def\d{\delta}        
\def\e{{\rm e}}           
\def\z{\zeta}              
\def\j{\eta}

\def\m{\mu}	           
\def\n{\nu}                
\def\x{\xi}              
                  
\def\p{\pi}        \def\P{\Pi}      
\def\r{\rho}

\def\y{\psi}

\def\yh{\hat{\y}}
\def\yhd{\hat{\y}^{\dag}}

\def\w{\omega}     
   
\def\br{\langle}
\def\ket{\rangle}
\def\ve{\vert}

\def\zbar{\bar{z}}

\def\eqref#1{(\ref{#1})}
\def\zexp{\e^{-{{B}\over 2}\ve z \ve^2 }}

\def\enorm{\e^{{2\over{B}}\del_{\eta }\del_{\bar{\z }}}}

\def\zpexp{\e^{-{{B}\over 2}\ve \zp \ve^2 }}

\def\zp{z^{\prime }}
\def\zbarp{\bar{\zp }}
\def\zzprint{\int d^2z d^2\zp }
\def\zzprexp{\zzprint \e^{-{{B}\over 2}(\vert z \vert^2 +\vert \zp \vert^2 )}}

\def\dt{\int dt}
\def\dx{\int d\vec x}
\def\dxx{\int d\vec x^{\prime } }
\def\dxi{\int d^2 \x \ \e^{-{{B}\over 2}{\vert \x \vert^2}}}

\def\Phx{\hat {\P^x }}
\def\Phy{\hat {\P^y }}

\def\Ph{\hat \P }
\def\Phd{\hat {\P^{\dag }}}
\def\Xh{\hat X}
\def\Yh{\hat Y}
\def\zh{\hat z}
\def\zbh{\hat{\zbar }}
\def\Zh{\hat Z}
\def\Zbh{\hat {\bar Z}}

\magnification=1200
{\bf  Appendix A}
\bigskip
In this appendix, we shall elaborate upon some results used 
extensively in the main text.
\bigskip
{\bf a.}
\bigskip
$$
\br \bar \z \ve \sharp f(\Zh ,\Zbh )\sharp \ve \eta \ket = 
\enorm \br \bar \z \ve : 
f(\Zh ,\Zbh ): \ve \eta \ket \eqno (A.1).
$$
This is shown as follows:
$$
f(\Zh ,\Zbh )=\int d\vec k f(\vec k)\e^{i{{\bar k}\over 2}\Zh + 
i{{k}\over 2}\Zbh } \eqno (A.2).
$$
Thus, by definition,
$$
\sharp f(\Zh ,\Zbh )\sharp = \int d\vec k \e^{i{{\bar k}\over 2}\Zh }
\e^{i{{k}\over 2}\Zbh }f(\vec k) \eqno (A.3).
$$
Using the result
$$
\e^{A} \e^{B} = \e^{B} \e^{A} \e^{[A,B]}
$$
with
$A\equiv i{{\bar k}\over 2}\Zh $, $B\equiv i{{k}\over 2}\Zbh $ and 
$[A,B]=-{{{\vec k}^2}\over {2B}}$, we get, from (A.3),
$$
\sharp f(\Zh ,\Zbh ) \sharp =\int d\vec k \e^{-{{{\vec k}^2}\over {2B}}}
\e^{i{{k}\over 2}\Zbh } \e^{i{{\bar k}\over 2}\Zh }f(\vec k)  
\eqno (A.4). 
$$
Thus, sandwiching (A.4) between coherent states, we get (A.1) directly.
\bigskip
{\bf b.}
\bigskip
Another expression that we repeatedly encounter is:
$$
A_{f,g}\equiv \br \bar z \ve \bigl[ \dxi : f(\Zh ,\Zbh ) : \ve 0,\xi \ket 
\br 0,\bar \xi \ve : g(\Zh ,\Zbh ) : \bigr] \ve z \ket \eqno (A.5).
$$
Since we know the action of $\Zh $ and $\Zbh $ on the coherent states,
we obtain from (A.5)
$$
A_{f,g}=\int d^2\xi \e^{-{{B}\over 2}\ve \xi \ve^2 + {{B}\over 2}\bar z
\xi + {{B}\over 2}\bar \xi z}f(\xi ,\bar z)g(z,\bar \xi ) \eqno (A.6).
$$
Shifting, $\xi = \j + z$ and $\bar \xi = \bar \j + \bar z$, we get
$$
A_{f,g}=\e^{{{B}\over 2}\ve z \ve^2 }\int d^2 \j \e^{-{{B}\over 2}
\ve \j \ve^2 }f(z+\j ,\bar z)g(z,\bar z + \bar \j ) \eqno (A.7),
$$
which may be expanded as
$$
A_{f,g}=\e^{{{B}\over 2}\ve z \ve^2 }[f(z,\bar z)g(z,\bar z)
+{2\over {B}}\del_z f\del_{\bar z}g + \cdots ] \eqno (A.8),
$$
where the dots indicate terms of $O({1\over {B^2}})$.
\bigskip
{\bf Appendix B}
\bigskip
In section 5, we have described the contributions to the effective
action for the fermionic spin, arising from the mixing of the
higher L.L. with the L.L.L. This mixing is brought about by the 
Coulomb term. In this appendix, we derive an expression for the
Coulomb term, projected onto the L.L. basis.

The bare Coulomb term may be expressed as 
$$
V(\hat{\vec r_1 } - \hat{\vec r_2 }) = \int {{d\vec k }\over {2\pi }}
\e^{i\vec k \cdot (\hat{\vec r_1 } - \hat{\vec r_2 })}{{e^2}\over{\ve \vec k \ve 
}}
\eqno (B.1)
$$
where $\vec r_{i} \equiv (x_{i} , y_{i} ).$

Thus, the Coulomb Lagrangian is 
$$\eqalignno{
L_c = -\int {{d\vec k }\over {4\pi }}{{e^2}\over{\ve \vec k \ve }}
\sum_{n_1,n_2,n_1^{\prime },n_2^{\prime }=0}^{\infty }
&\int \prod_{i=1}^{2} d^2\x_{i} d^2\x_{i}^{\prime }
\e^{-{{B}\over 2} (\ve \x_{i} \ve^2 + \ve \x_{i}^{\prime }
\ve^2 )}W_{n_1,n_1^{\prime },n_2,n_2^{\prime }}(\vec k) \cr 
&\yhd_{n_1}(\x_1)\yhd_{n_2}(\x_2)\yh_{n_2^{\prime }}
(\bar{\x_2^{\prime }})
\yh_{n_1^{\prime }}(\bar{\x_1^{\prime }}) & (B.2)\cr 
}
$$
where 
$$
W_{n_1,n_1^{\prime },n_2,n_2^{\prime }}(\vec k) 
\equiv \br n_1,\bar{\x }_1 \ve \br n_2,\bar {\x }_2 \ve 
\e^{-i\vec k \cdot (\hat{\vec r_1}-\hat{\vec r_2})}\ve n_2^{\prime },
\x_2^{\prime }\ket \ve n_1^{\prime },\x_1^{\prime }\ket 
\eqno (B.3).
$$
Expressing the coordinate operators in terms of $\Zh , \Zbh , \Ph , \Phd $
we get
$$
W_{n_1,n_1^{\prime },n_2,n_2^{\prime }}(\vec k) = 
\e^{-i{{k}\over 2}(\bar{\x_1}-\bar{\x_2})-i{{\bar{k}}\over 2}
(\x_1^{\prime }-\x_2^{\prime })}\e^{{{B}\over 2}(\bar{\x_1}\x_1^{\prime }+
\bar{\x_2}\x_2^{\prime })}M_{n_1,n_1^{\prime }}(\vec k)M_{n_2,n_2^{\prime }}
(-\vec k)
\eqno (B.4).
$$
Here,
$$
M_{n,n^{\prime }}(\vec k)\equiv \br n \ve \e^{{{k}\over{2B}}\Ph }
\e^{-{{\bar{k}}\over{2B}}\Phd }\ve n^{\prime }\ket 
\eqno (B.5),
$$
with $k\equiv k_{x}+ik_{y}$ and $\bar{k}\equiv k_{x}-ik_{y}$.

Upon simplification,
$$
M_{n,n^{\prime }}(\vec k)={1\over{\sqrt{n!n^{\prime }!}}}
(-{{\bar{k}}\over{\sqrt{2B}}})^{n-n^{\prime }}
\sum_{l=max(0,n^{\prime }-n)}^{\infty }{{(n+l)!}\over{l!(n-n^{\prime }+l)!}}
(-{{k^2}\over{2B}})^{l}
\eqno (B.6).
$$
Using $\e^{{{B}\over 2}(\bar{\x }_2 + {{i\bar k}\over {B}} )\x_2^{\prime }}
= \br \bar{\x }_2 + {{i\bar k}\over {B}}\ve \x_2^{\prime } \ket  $,
we get,
$$
\e^{-{{B}\over 2}\ve \x_2 \ve^2 + i{{\bar k}\over 2}\bar{\x }_2 }
\int d^2\x_2^{\prime }\ \e^{-{{B}\over 2}\ve \x_2^{\prime } \ve^2 }
\e^{{{B}\over 2}(\bar{\x }_2+{{i\bar k}\over{B}})\x_2^{\prime }}
\yh_{n_2^{\prime }}(\bar {\x }_2^{\prime })
= \e^{{{k^2}\over {2B}}-{{B}\over 2}\ve \x_2 \ve^2 + i{{\bar k}\over 2}\x_2 
+ i{{k}\over 2}\bar {\x }_2}\yh_{n_2^{\prime }}(\bar {\x }_2) .
$$
Using this to integrate the primed variables out, we obtain, from (B.2),
$$\eqalignno{ 
L_c = -e^2 \int {{d\vec k }\over {4\pi }}{{\e^{{{k^2}\over{B}}}}
\over{\ve \vec k \ve }}
&\sum_{n_1,n_2,n_1^{\prime },n_2^{\prime }=0}^{\infty }
M_{n_1,n_1^{\prime }}(\vec k)M_{n_2,n_2^{\prime }}(-\vec k)
\int \prod_{i=1}^{2}d^2\x_{i}\e^{-{{B}\over 2}\ve \x_{i} \ve^2 }
\e^{-i{{k}\over 2}(\bar{\x_1}-\bar{\x_2})-i{{\bar{k}}\over 2}
(\x_1-\x_2 )}\cr 
& \yhd_{n_1}(\x_1)\yhd_{n_2}(\x_2)\yh_{n_2^{\prime }}
(\bar{\x_2})\yh_{n_1^{\prime }}(\bar{\x_1})
& (B.7).\cr 
}$$
This is the Coulomb Lagrangian, projected onto the L.L. basis. The
term with all the L.L. indices set to zero is the Coulomb interaction
projected onto the L.L.L. and has been used extensively in the 
earlier sections of the article. 

>From (B.6), it is easily seen that,
$$
M_{n,n}(\vec k)=M_{n,n}(-\vec k)={1\over{n!}}{{\del^n }
\over{\del \r^{n}}}\ve_{\r =-{{k^2}\over {2B}}}(\r^{n}\e^{\r })
=\e^{-{{k^2}\over{2B}}}+ O(1/B)
\eqno (B.8)
$$
for all $n\ge 0$.
Again, for $n\neq 0$, we have
$$
M_{0,n}(\vec k)={1\over{\sqrt{n!}}}({{k}\over{\sqrt{2B}}})^{n}
\e^{-{{k^2}\over{2B}}};\ 
M_{n,0}(\vec k)={1\over{\sqrt{n!}}}({{-\bar{k}}\over{\sqrt{2B}}})^{n}
\e^{-{{k^2}\over{2B}}}
\eqno (B.9).
$$
This shows quite clearly that mixing between different Landau levels
is suppressed by powers of ${1\over{\sqrt{B}}}$ and hence, if we
look at the mixing of higher L.L. with the L.L.L., it would suffice
to consider only $n=1$.

\bigskip
{\bf Appendix C}
\bigskip

In this appendix we make contact with various representations of the
Goldstone fields found in the literature.

Let us start with $M=U\sigma^3 U^{\dagger}$ and recall that
$$
 S^{\dagger}=S
 \quad\quad ,\quad
 trS=0 \quad\quad , \quad S^2=1
 \eqno (C.1)
$$
Then we can write
$$
S= \vec \sigma \vec m
 \quad\quad ,\quad
{\vec m}^2=1
\eqno (C.2)
$$
Since $S\rightarrow g S g^{-1}$ , $\vec m$ transforms in the vector
representation of $SU(2)$. In terms of $\vec m$, for instance, the
kinetic energy term reads
$$
tr(\partial_{z} S\partial_{\bar z} S)=2
\partial_{z} \vec m\partial_{\bar z} \vec m
\eqno (C.3)
$$
It is also very easy to check that
$$
tr
(\partial_{z}U\sigma^3\partial_{\bar z}U^{\dagger} -
\partial_{\bar z}U\sigma^3\partial_{z}U^{\dagger})
=-{i\over 8}
\epsilon^{\alpha\beta}
tr(S
\partial_{\alpha}S
\partial_{\beta}S)
={1\over 4}
\epsilon^{\alpha\beta}
\epsilon_{ijk} m^{i}
\partial_{\alpha}m^{j}
\partial_{\beta}m^{k}
\eqno (C.4)
$$
$(\alpha , \beta = x , y )$
 which is a more conventional form of the
topological density.

Next, consider the term with the time derivative
$$
T=tr(\sigma_3 U^{\dagger}i\partial_0 U)
\eqno (C.5)
$$
As mentioned before this term cannot be written in terms of $S$ or $\vec
m$ in a local form. However, we can write it locally in $S$ or $\vec m$
if we introduce an extra dimension in the following way. We interpolate
smoothly the Goldstone fields
$\pi^{\alpha}(x)\rightarrow
\pi^{\alpha}(x,\lambda)
\quad , \lambda\in [0,1]$ in such a way that
$
\pi^{\alpha}(x,1)=
\pi^{\alpha}(x)$
and
$\pi^{\alpha}(x,0)=0 $
. Then it is very easy to check that
$$
T=-{i\over 4}
\int_0^1d\lambda
\epsilon^{\alpha\beta}
tr(S
\partial_{\alpha}S
\partial_{\beta}S)
=
{1\over 2}
\int_0^1d\lambda
\epsilon^{\alpha\beta}
\epsilon_{ijk} m^{i}
\partial_{\alpha}m^{j}
\partial_{\beta}m^{k}
\eqno (C.6)
$$
$(\alpha , \beta = 0 , \lambda )$.
To our knowledge, the last expresion for $T$ was first written down in [15], whereas 
both the second last expression, which resembles a two dimensional Wess-Zumino term, and
(C.5) are original.
  $T$ is usually found with a rather different aspect, namely,
$$
T=\vec A (\vec m)\partial_0 \vec m
 \quad\quad ,\quad
\epsilon_{ikj}
{\partial A^{k}\over \partial m^{j}}=
m^{i}
\eqno (C.7)
$$
It is simple to check that (C.7) and (C.6) lead to the same equation of 
motion.

\end

Finally, if we want to write the lagrangian in terms of properly
normalized fields, we just have to expand
$$
U=e^{i\pi_{\alpha}\sigma^{\alpha}\over f_{\pi}}=1+
i{\pi_{\alpha}\sigma^{\alpha}\over f_{\pi}}+ \cdots
 \quad\quad ,\quad
\alpha=1,2
\eqno (C.9)
$$
 The lagrangian to quadratic order reads
$$
  \eqno (C.10)$$
^Z